\documentclass[apj]{emulateapj}

\def\<#1>{\langle#1\rangle}
\def\pcite#1{[ref]}

\slugcomment{New manuscript: \today}
 
\shortauthors{Chakrabarty $\&$ Raychaudhury}
\shorttitle{Dark Matter Distribution in NGC 4636}
 
\begin{document}    

\title{The Distribution of Dark Matter in the Halo of the early-type galaxy NGC~4636}
\author{Dalia Chakrabarty\altaffilmark{1} and
  Somak Raychaudhury\altaffilmark{2}}
\altaffiltext{1}{School of Physics \& Astronomy, University of Nottingham, 
Nottingham NG7 2RD, U.K.; 
dalia.chakrabarty@nottingham.ac.uk}
\altaffiltext{2}{School of Physics \& Astronomy, 
    University of Birmingham,
    Birmingham B15 2TT, UK; somak@star.sr.bham.ac.uk}

\begin{abstract}
\noindent
We present the density structure of dark matter in the outer parts (to
about 7 effective radii) of the elliptical galaxy NGC 4636, from the
radial velocities of 174 globular clusters, using the non-parametric,
inverse algorithm CHASSIS. We find the galaxy to be rich in dark
matter, with $R$-band mass-to-light ($M/L$) ratios rising to about 30,
at nearly 4$R_e$; the $K$-band $M/L$ at about 3$R_e$ is found to be
nearly 10. The result does not depend on applying the method to the
red and blue globular clusters separately. This estimate of $M/L$ is
higher than the previous analysis from the same kinematic data. We
also find that the dark matter distribution is highly concentrated
towards the inner halo.

\end{abstract}


\section{Introduction}
\label{sec:intro}
\noindent
The overwhelming presence of dark matter in the outskirts of disk
galaxies has been invoked to explain flat rotation curves out to
several times the optical dimensions of these systems.
\citep[e.g.,][]{rob69,fg79}. However, the inferred dark matter content
of cores of galaxies, in many cases, seems to fall well short of the
standard model predictions \citep[e.g.,][]{gentile04,fswb08}. In the
same spirit, intriguing studies of dark haloes of early-type galaxies,
using planetary nebulae as tracers of the distribution of matter, seem
to indicate the under-abundance of dark matter
\citep[e.g.,][]{roman03}. This view has been challenged, however, on
grounds that the subtleties in the very processing of the kinematic
data can lead to spurious answers \citep{dekel05, sambhus06,
douglas07}. In particular, the mass-anisotropy degeneracy is a
difficult issue to resolve, given the limited breadth of available
kinematic measurements.

In spite of this shortcoming, it is highly appealing to use kinematic
information in verifying that the dark matter distribution of
elliptical galaxies is in concordance with with the predictions of
cosmological simulations, performed within the $\Lambda$CDM paradigm.
Globular clusters provide an independent source of test particles,
since their formation histories are different from those of planetary
nebulae, and the nature and distribution of their orbits would provide
independent information about the distribution of the underlying dark
matter. Mass modeling based on globular cluster dynamics has been
tried out on nearby early-type galaxies
\citep{cote03,bridges06,schuberth06,woodley07,hwang08,richtler08}, and
in many cases there have been appreciable discrepancies between these
results and the models obtained from planetary nebulae. Such endeavors
highlight the necessity of not only studying a large number of systems
for both sets of tracers, but also investigating the dependence on
models, as well as trying out different methods of mass
reconstruction, in particular those that do not explicitly depend on
the characterization of models.

NGC~4636 is an example of an early-type galaxy for which the different
modes of mass modeling have led to a large discrepancy in the
recovered mass, even at the effective radius
($R_e$=101$^{\prime\prime}$.7, Schuberth et. al 2006). NGC 4636 lies
near the southern border of the Virgo cluster and is relatively
isolated. It has a radial velocity similar to that of the Virgo
cluster, even though it is about 3~Mpc from the center of the cluster.
Furthermore, the galaxy lies at the center of a poor group
\citep{op04,miles04,miles06}, possibly falling into the cluster. Its
unusual properties have attracted detailed multi-wavelength research
for several decades.  For instance, NGC~4636 is found to be very
bright in X-rays ($L_X\sim10^{41}$ergs/s), with unusual features in
the hot ISM \citep{forman85, matsushita98, jones02,
loewenstein03,ewan05}.

NGC~4636 has an anomalously large specific frequency of globular
clusters \citep{dirsch05}, comparable to that of central galaxies of
far richer systems, like that of NGC~1399 in Fornax \citep{dirsch03}.
\citet{dirsch05} provide details of several thousand globular cluster
candidates around this galaxy, of which radial velocities of 174 GCs
were presented in \citet{schuberth06}, where a dynamical analysis of
this data, on the basis of Jeans Equation, was also performed, for
assumed values of the stellar mass to light ratio ($M/L$) and
anisotropy. Naturally, the recovered mass profile is the projected
one, which only provides a lower limit to the actual mass
distribution. In this approach, the total mass density profile needs
to be parameterized (as an NFW model, say) and constraints are
recovered for these characteristic parameters. Thus, this method is
not well-suited for the recovery of the distribution of total
mass. Further, the estimation of the density parameters is awkwardly
sensitive to the choice of the stellar $M/L$ that is used to scale the
luminosity density profile: \citet{schuberth06} use $M/L$ values
obtained by \citet{bell03}, \cite{kronawitter} and
\citet{loewenstein03}, from various samples of elliptical galaxies.
There is of course no reason why this ratio has
to be a constant and not vary with radius in the radial range covered
by the kinematic data (projected radius $R\leq 30$~kpc).

According to \citet{schuberth06}, the range of mass profiles of the
halo of NGC~4636, corresponding to an assumed NFW density \citep{nfw},
do indeed straddle the distribution indicated by \citet{matsushita98},
though the profile obtained by \citet{loewenstein03} is found to be
too massive, even for the \citet{schuberth06} model that corresponds
to the highest dark matter content within 30~kpc.

Clearly, the use of different methods, with varying model assumptions,
on different data sets, disagree in their attempts to find the
underlying dark matter distribution within the same volume. Here, we
take an approach which does not assume a parameterized model. We use
the algorithm CHASSIS \citep{dalpras} to analyze the available
kinematics of the same sample of 174 GCs that is used in the Jeans
Equation approach of \citet{dirsch05}. This algorithm has been
calibrated against the N-body realization of two star clusters
\citep{chakrabartyzwart}, and applied to estimate the central mass
structure of the Galaxy \citep{dalpras} and the GC M15
\citep{chakrabartym15}. It is also used to estimate the dark matter
content in NGC~3379, using planetary nebulae kinematics (Chakrabarty
2008, in preparation).

The basic formalism of CHASSIS is discussed in the following
section. Section~3 describes the observational details of the GCs used
in this analysis, and in Section~4, we present the results obtained
from CHASSIS for the $M/L$ distribution in the outer parts of NGC
4636, and compare our results with previous work.  
Section~5 summarizes the implications of our results in the determination
of the content of dark matter in the outer halo of NGC~4636.

\section{The CHASSIS Algorithm}
\label{sec:algorithm}
\noindent
CHASSIS works under the assumption that the input kinematic data is
drawn from an equilibrium phase space distribution function that is
isotropic in nature. Also, the system geometry is assumed to be
spherical. The algorithm produces a pair of functions: the equilibrium
phase space density and the (total) mass density, that best describe
the observed data. These two characteristic functions are sought
simultaneously, using a maximum likelihood approach that employs a
sophisticated optimizer -- the Metropolis algorithm. The inputs to the
algorithm are the position on the plane of the sky, and at least one
velocity component (usually the radial velocity $v_z$) of individual
GCs in the system.

As in all recursive algorithms, CHASSIS too requires initial guesses
(or seeds) for the answers it seeks, namely the distribution function
and mass density profile (represented in the code as histograms over
energy and radius, respectively). The final answer should be
insensitive to the choice of this guess; robustness checks are carried
out to confirm this with the kinematic data used.

During any run, at the end of the first step, the seeds for the mass
density and the distribution function are slightly tweaked. Likewise,
at the end of subsequent step, the profiles are modified, (both in
shape and overall amplitude), over their previous forms, subject to
the constraints of monotonicity and positivity. This is carried on
until the global maximum in the likelihood function is identified.

The $\pm$1-$\sigma$ spread in the sample of the density and
distribution functions, that correspond to the neighborhood of this
global maximum, readily provides the 1-$\sigma$ error bars on the
recovered profiles. Quantities that are estimated from the recovered
profiles, such as the enclosed mass profile and the velocity
dispersion profile, bear the signature of this extent of error. It
may be noted that these errors stem from the uncertainties in
identifying the global maximum in the likelihood function, and are
essentially different from the observational errors. The errors in the
velocity measurements are incorporated into the analysis by convolving
the projected distribution function with the distribution of the
observational errors, (assumed Gaussian). Further details about
this algorithm can be found in \citet{dalpras}.

CHASSIS works by projecting the distribution function in each step, at
the current choice of the potential (calculated from the current
choice of the mass density), into the space of observables. The
product of the resulting projected distribution functions,
corresponding to all the ($r_p$, $v_z$) pairs in the data set, defines
the likelihood function.

\begin{figure*}
\plotone{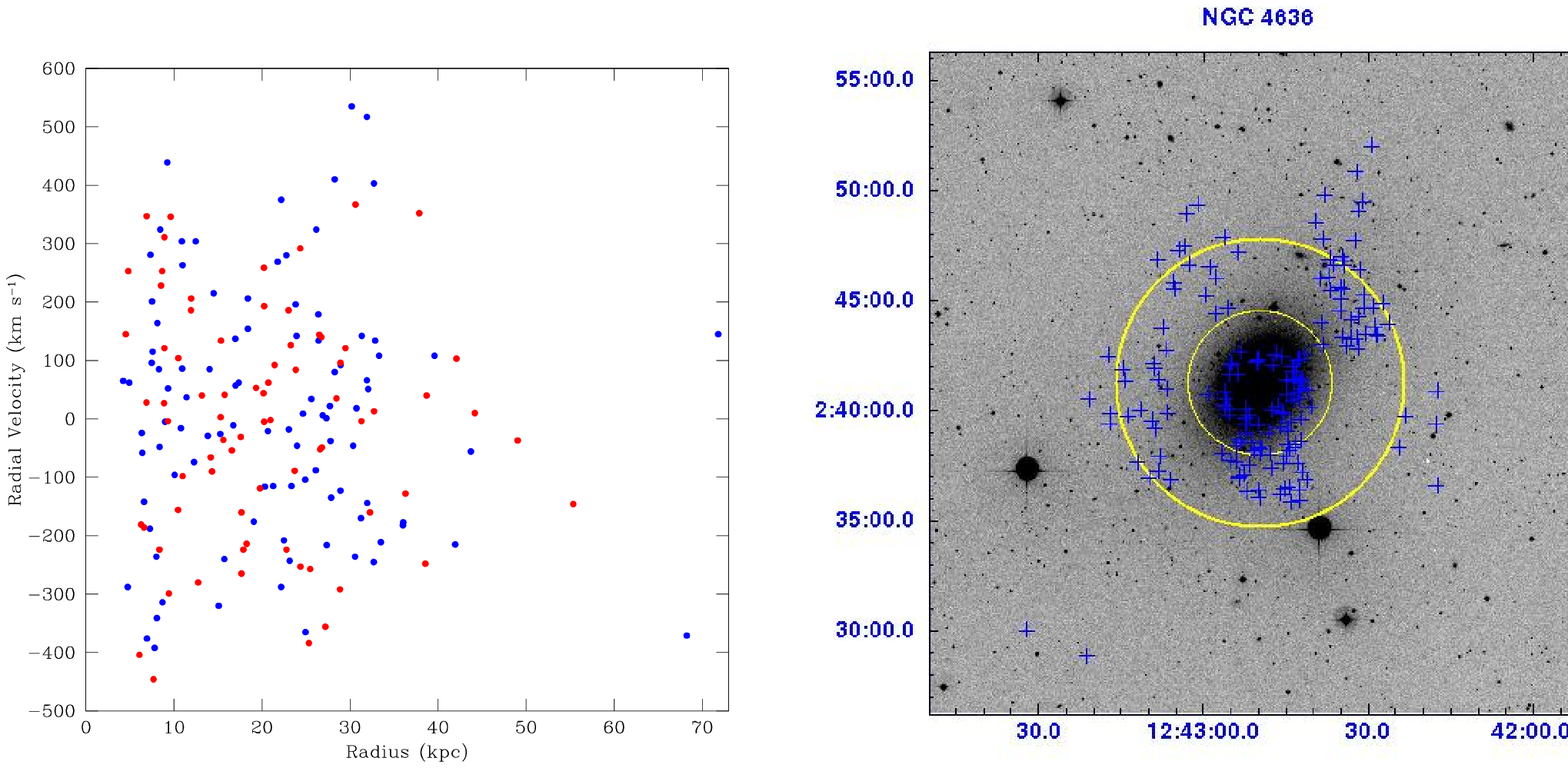}
\caption{Left: Measured radial velocities of the globular clusters of
NGC 4636 used in this analysis, as a function of projected distance
from the center of the galaxy.  The red points represent the 76 red
GCs ($C\!-\! R$ color $\!>\! 1.55$), while the blue points the 98 blue
GCs ($C\!-\!R \!\le\! 1.55$).  Right: The distribution of these GCs
(blue crosses) on the sky, superposed on a optical DSS image of NGC
4636.  The two yellow circles correspond to radii of 15 and 30 kpc
respectively.}
\label{fig:data}
\end{figure*}

\section{Data}
The positions and radial velocities ($v_z$) for the 174 globular
clusters that are used in this study, (adopted from \citet{dirsch05}),
are shown in Fig.~\ref{fig:data}. Based on surface brightness
fluctuations, \citet{tonry01} quotes a distance to NGC~4636 to be 14.7
Mpc, while \citet{dirsch05} quote a distance of 17.7~Mpc based on the
peak of the GC luminosity function. Both measures have considerable
uncertainty, so we adopt a distance of 16 Mpc to NCG 4636, which
translates to 1$^{\prime\prime}\equiv$77.5 pc, approximately.

As has now been found in a wide range of early-type galaxies, the
colors of the GCs form a bimodal distribution, with the redder, more
metal-rich GCs more likely to harbor low-mass X-ray binaries
\citep[e.g.][]{jordan04,posson06,kz07,woodley08}, there being
significant differences in key structural as well as chemical
properties in the two populations \citep[e.g.][]{jordan07}. According
to the photometry of \citet{dirsch05}, the blue and red GCs of
NGC~4636 can be characterized as $C-R$ color being greater than or
less than 1.55 respectively. We show this classification in the left
panel of Figure~\ref{fig:data}. Another way of classifying the data
at hand is along the lines of the magnitude of the measured velocity
errors; any GC with a measurement error exceeding 35 kms$^{-1}$ is
assigned to one kinematic class while those with higher velocity
errors for the other group. We carry out runs with all GCs, and
separately with the sub-samples that characterize each of these
photometric and kinematic classes.

The right panel of Fig.~\ref{fig:data} shows the spatial distribution
of the globular clusters used in this study. This is a small fraction
of the GC candidates of \citet{dirsch05}, and represents the clusters
that \citet{schuberth06} report as belonging to the galaxy, on the
basis of spectroscopic data. As is apparent from this picture, the
number of observed GCs decreases rapidly beyond about 30 kpc, and
beyond 50 kpc, there are only 2 velocity measurements available. This
radial profile of the measured data set is important in determining
the choice of the radial binning adopted in this analysis.

For the application of CHASSIS, the radial bin width needs to be such
that the left edge of the innermost bin must not exceed the smallest
projected radius at which the radial velocity data is
available. Working with too small bin widths would lead to bins that
contain no velocity information at small radii.  On the contrary,
adopting too large a bin width can lead to spurious density profiles,
especially nearer to the center of the system, where the gradient in
density is higher than in the outer parts. Given these constraints, we
experimented with bin sizes and found that a choice of 2 kpc is
adequate for this data set. This value allows us to span radial
distances over a range of 4.2 kpc to about 55 kpc. However, the
results recovered by CHASSIS are not crucially sensitive to the exact
choice of the bin width; this feature of the code will be demonstrated
below with results from experiments done with assorted bin width
values.

\begin{figure*}
\plotone{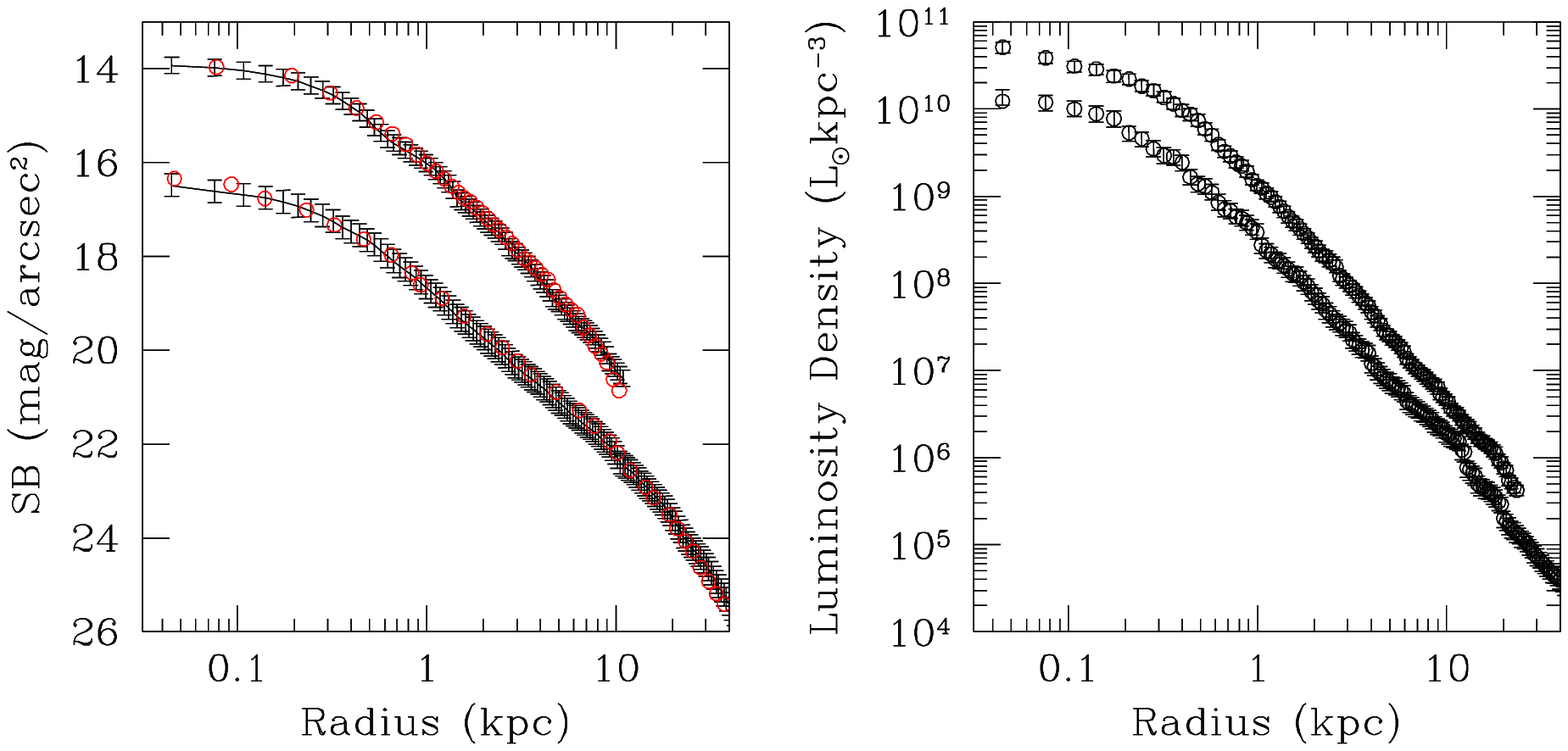}
\caption{The $R$-band (lower curve; from Dirsch et. al 2005) and
$K$-band (upper curve; 2MASS, \citet{jarrett03}) surface brightness
profiles of NGC 4636 along the major axis, are shown in red on the
left. The deprojected luminosity density profiles (upper curve for
$K$-band and lower curve for $R$-band) are shown on the right and the
projections of the same have been overlaid in black over the
brightness profiles, in the left panel. The deprojection was carried
out assuming that the intrinsic geometry of the tracer population is
spherical.}
\label{fig:iden}
\end{figure*}

The 3-D total mass density that is recovered by CHASSIS is then used
to calculate the enclosed mass profile which is subsequently compared
to the cumulative light profile, in order to extract the mass-to-light
ratio ($M/L$) profile. In this context, we use the $K$-band surface
brightness profile of NGC~4636, obtained from the photometry of the
2MASS large galaxy Atlas \citep{jarrett03}, kindly provided to us by
Tom Jarrett. This is shown in red in the left panel of
Fig.~\ref{fig:iden}). We deproject this under the assumption of
spherical symmetry (right panel of Fig.~\ref{fig:iden}) with a
non-parametric deprojection code DOPING \citep{doping}. We also use
the $R$-band photometry from \cite{dirsch05} to estimate the $M/L$
profile in the $R$-band.

\begin{figure*}
\plotone{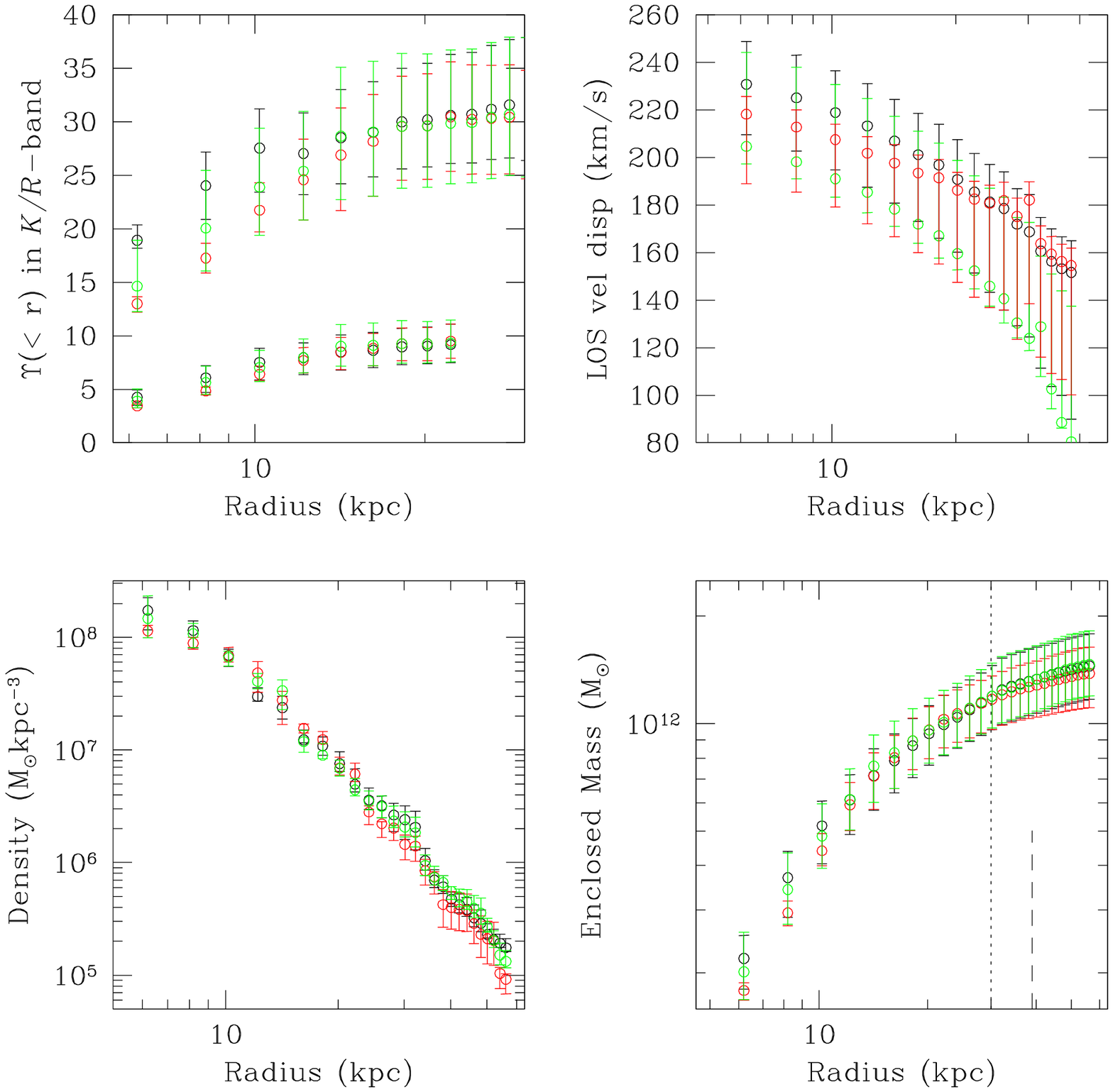}
\caption{{\it Lower left:} The recovered 3-D total (dark+luminous)
density profiles of NGC 4636, to about 7 times the effective radius
(which is 101$^{\prime\prime}.7 \approx 7.8$~kpc), from three
different runs, that were performed with distinct forms of the initial
guesses (seeds) for the DF and mass density distributions. Results for
RUN~I are in black, RUN~II in red and RUN~III in green (details in
\S4).  {\it Lower right:} Enclosed mass profiles from the density
distributions recovered from these runs. The dotted line in black
represents 30 kpc, the radius at which the mass predicted by CHASSIS
is compared to the mass estimates from other work. The dashed line
represents the 5$R_e$ mark.  {\it Upper left:} The $K$-band (lower
curve) and $R$-band (upper curve) mass-to-light ratios in the outer
parts of NGC 4636. Note that the $M/L$ distributions in this figure do
not include the uncertainties involved in the deprojection of the
observed brightness profile into the intrinsic luminosity profile, in
order to represent the extent of the errors in the analysis due to
CHASSIS alone; allowing for these errors would widen the range of
$\Upsilon_K$ at about 24 kpc (edge of the available photometry) from
about 9--12 to about 8--14. The quality of the $R$-band photometry
implied truncating the $M/L_R$ estimate to within 30~kpc even though
mass estimates are available to larger distances.  {\it Upper right:}
The velocity dispersion along the line-of-sight, estimated from the
isotropic equilibrium distribution function recovered by CHASSIS, from
the runs RUN~I, RUN~II and RUN~III.}
\label{fig:robust}
\end{figure*}

\section{Results}
\label{sec:result}
\begin{figure}
\plotone{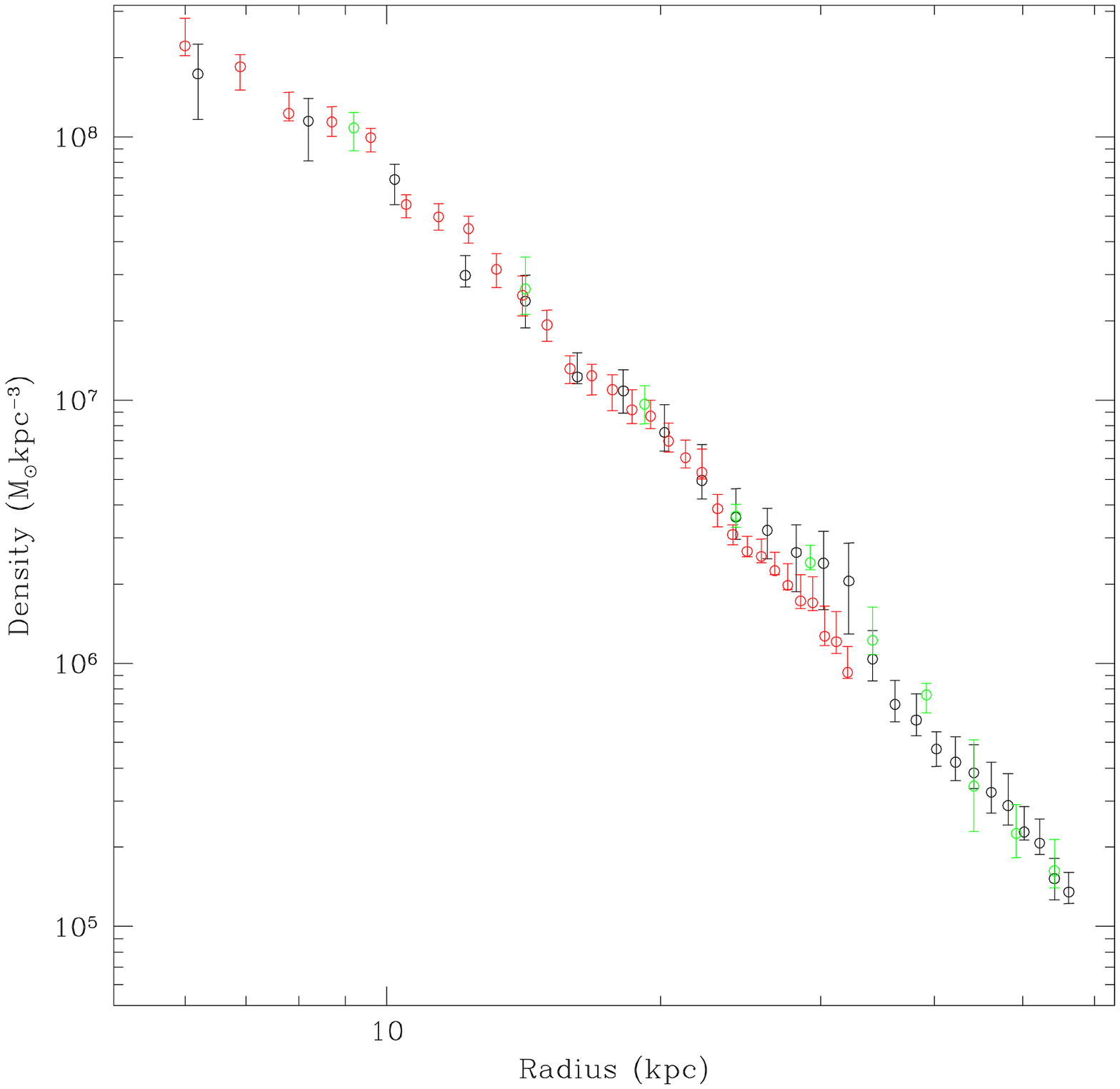}
\caption{This shows the concordance (within $\pm$1-$\sigma$ error
bars) of density profiles recovered from runs performed with initial
conditions corresponding to RUN~I and bin widths of 2 kpc (in black), 1
kpc (in red) and 5 kpc (in green). The unavailability of kinematic
information in 1 kpc sized bins, beyond about 30 kpc, limits the
radial range over which the density can be sought in this case.}
\label{fig:bin_size}
\end{figure}

\noindent
To begin with, the insensitivity of the results to the choice of the
first guess for the distribution function needs to be established. For
this purpose, we undertake to parameterize the seed for the density by
an NFW-like profile \citep{nfw}:
\begin{equation}
\rho(r) = \frac{\rho_0}{(r/r_c)^{\alpha}(1+r/r_c)^2}.
\label{eqn:guess_rho}
\end{equation}
The phase space distribution function is either held as a power-law
(with a power-law index of $\beta$) or as an exponential of the
effective energy ($\epsilon$). Thus, the seed for the density
distribution is characterized by a total of 3 parameters, namely
$\rho_0$, $r_c$, $\alpha$, out of which, the amplitude or the central
density parameter $\rho_0$ is found to have no effect whatsoever on
the result. Three different seeds characterized by the form of the DF
and values of $r_c$, $\alpha$, were implemented in three distinct runs:
\begin{itemize}
\item RUN~I -- the seed distribution function is set to
$\exp(\epsilon)$, $r_c$=50 kpc and $\alpha$=2.8; kinematic data of all 174 GCs are used.
\item RUN~II -- the seed distribution function 
$\propto \epsilon^2$, $r_c$=5 kpc and $\alpha$=1.8; all GCs are used.
\item RUN~III -- the seed distribution function is 
$\propto \epsilon$, $r_c$=15 kpc and $\alpha$=2.3; again, the whole sample is used.
\end{itemize}
The density profiles resulting from these three runs are shown in the
lower left panel of Fig.~\ref{fig:robust}.  As indicated by the
results from the different runs, the recovered density profiles
overlap within the error bars, at almost all radii. This lends
confidence in the functionality of CHASSIS. The estimated density
profiles are then used to calculate the enclosed mass distributions
from which (lower right in Fig.~\ref{fig:robust}), the radial
distributions of $M/L$ are recovered in the $R$ and $K$ bands (upper
left Fig.~\ref{fig:robust}). Since the $K$-band photometry is
available to about 23.5 kpc, the run of $M/L$ is also displayed till
this radius, while the $R$-band $M/L$ is limited to 30~kpc. As
expected from a typical color of an early-type galaxy, the lower limit
on the ratio between the $R$-band and $K$-band $M/L$ values, at 20
kpc, is about 2.5. The DF recovered from three of the runs is used to
calculate the velocity dispersion along the line-of-sight. These have
been represented in the upper right in Fig.~\ref{fig:robust}.  (It is
worth pointing out here that in the analysis of Chandra and XMM-Newton
observations \citep{ewan05}, it has been found that between
25--30~kpc, there is an abrupt transition between the hot interstellar
medium of the galaxy to the intergalactic medium of the surrounding
group). The robustness of the algorithm to the choice of bin width is
displayed in Figure~\ref{fig:bin_size}.

\begin{figure*}
\centerline{
\includegraphics[width=0.95\hsize]{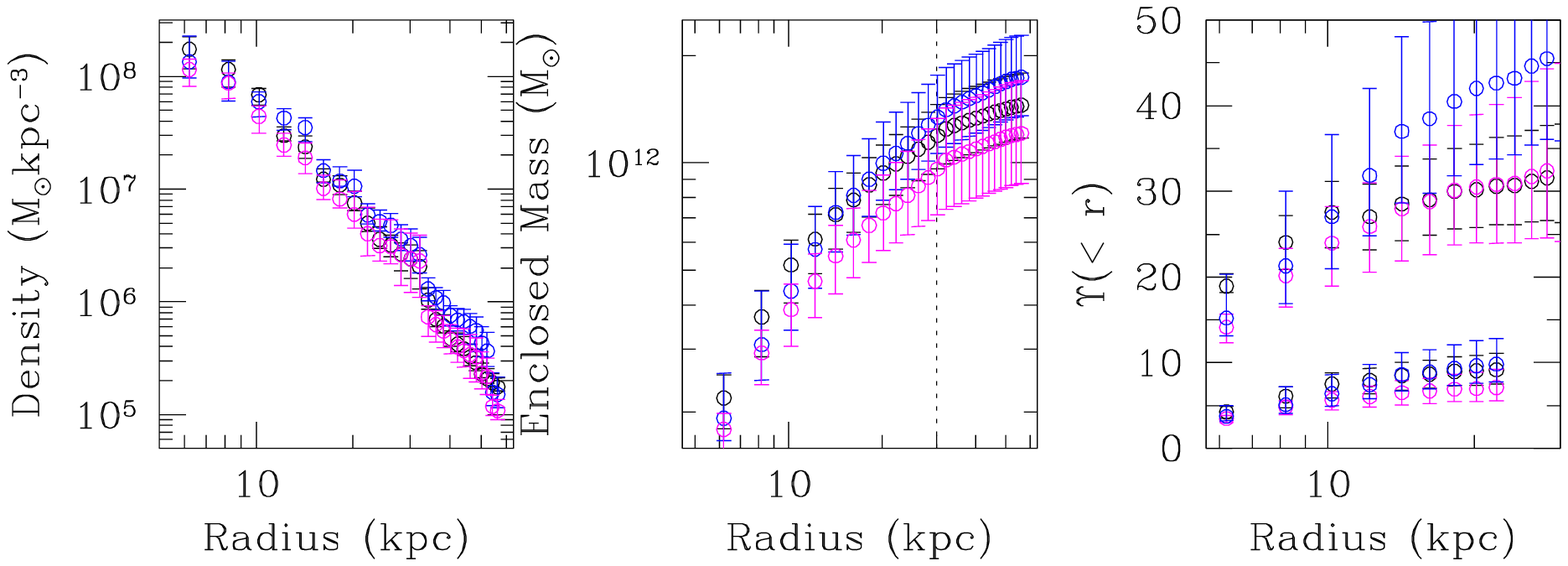}}
\centerline{
\includegraphics[width=0.95\hsize]{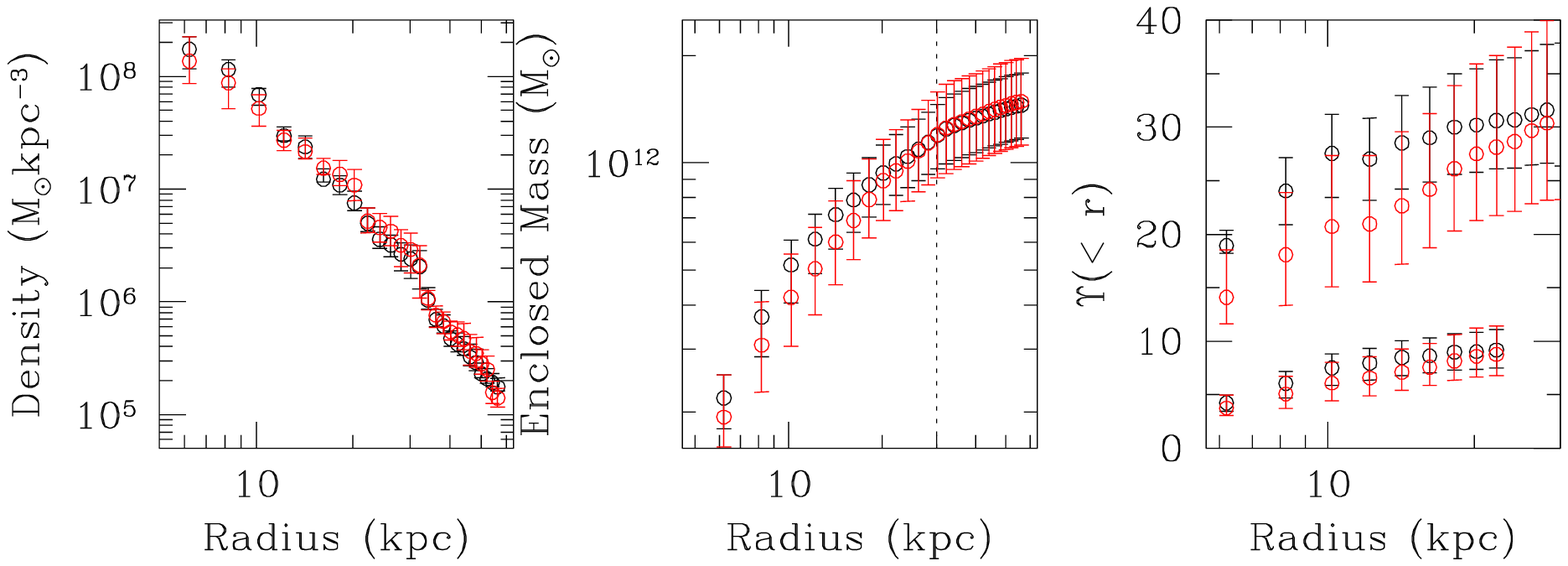}
}
\caption{
{\it Top Panel:} Total mass density, enclosed mass and  $M/L$ in the
$R$ (upper curve) and $K$-band (lower curve), from kinematic data of
all clusters (RUN~I, in black), red GCs (RUN~IV, in pink) and blue GCs
(RUN~V, in blue). 
{\it Bottom Panel:} Total mass density, enclosed mass and $M/L$ in the
$R$ (upper curve) and $K$-bands (lower curve) from kinematic data of
all clusters (RUN~I, in black) and the GCs with the relatively better
kinematic accuracy (RUN~VI, in red).}
\label{fig:colour}
\end{figure*}

To confirm that the results of RUN~I, RUN~II and RUN~III are not
artifacts of the details of the sampling of the tracer population, we
performed individual runs with sub-samples that correspond to
different GC color and (radial velocity) measurement accuracy
classes. These runs are described below:
\begin{itemize}
\item RUN~IV - the seed distribution function is set to the same as
for RUN~I; kinematic data of the 76 red GCs are used.
\item RUN~V - the seed is as in RUN~I; the  98 blue GCs are used.
\item RUN~VI - again, the seed is identical to that used in RUN~I;
implemented velocity data are that of the 121 GCs (red {\it and} blue)
that have velocity errors lower than a cutoff of 35 kms$^{-1}$
\citep{schuberth06}.
\end{itemize}
These three runs are carried out with bin widths of 2~kpc. The size of
the error bars on the recovered DF and density distributions will be
higher for these runs than for RUNS~I, II and III, which employ the
whole sample, i.e. a greater number of data points. The results are
shown in Fig.~\ref{fig:colour}. 

It is notable that the division of the whole sample of the GCs by
color did not yield significantly different total mass densities;
there is indeed a trend for the bluer GCs in the sample to be on the
higher side of the mass of the redder GCs, but as is apparent from
Fig.~\ref{fig:colour}, the distinction in the recovered mass profile
is not strong enough to be deemed significant, at $\pm$1-$\sigma$
level. The effect of division by kinematic accuracy is even less
potent.

Thus, it appears that the $M/L_K$ distribution within the first three
effective radii, can at most be about 12 in the $K$-band, when the
errors of the deprojection, as performed by DOPING are ignored; when
the errors of the deprojection are included, the range of the $M/L$
value recovered at about 23.5 kpc ($\approx R_e$) is found to widen to
about 8 to 14.

\begin{figure*}
\centerline{
\includegraphics[width=1.05\hsize]{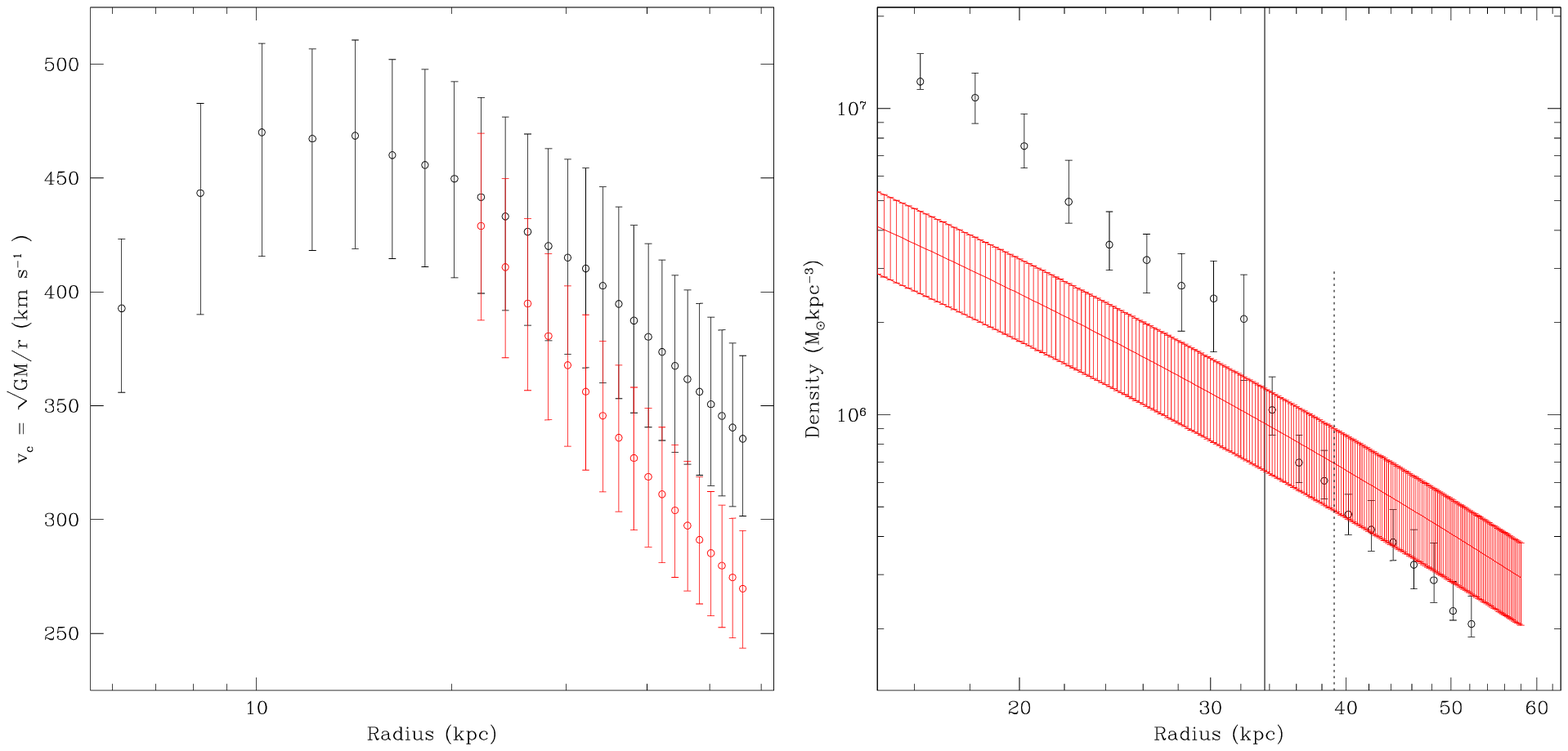}}
\caption{The left panel shows the circular velocity curve of NGC 4636,
as recovered from RUN~I. The function in red represents the Keplerian
fall-off with radius, normalized by $\sqrt{GM_{\rm 3R_e}}$, where
$M_{\rm 3R_e}$ is the mass found to be enclosed within 3$R_e$. The
errors on the red profile are the $\pm$1-$\sigma$ errors in the value
of $M_{\rm 3R_e}$, as indicated by the algorithm. The right panel shows
an NFW fit (in red) to the total mass density recovered from RUN~I (in
black). The errors of the fitting procedure define the error band
shown in red. The solid line represents the inner edge of the radial
range over which the fit is sought. The dotted line indicates
5$R_e$. }
\label{fig:vc}
\end{figure*}

\subsection{Effect of Assumptions on Mass Recovery}
\noindent
Within CHASSIS, the effect of imposing isotropy in velocity space,
where anisotropy prevails, is to recover spuriously high mass density
values \citep{chakrabartym15}. Though in general it is difficult to
quantify the extent of this bias since it depends on the structure of
the unknown distribution function, in the case of NGC~4636, we can
surmise that the effect of anisotropy is not significant at the
1-$\sigma$ level. We infer this using the following result that
\citet{schuberth06} report: it is the red sub-sample of the observed
GCs that were found to exhibit ``significant rotation'' in contrast to
the blue GCs which do not show a ``significant signal''.  However,
from Figure~\ref{fig:colour} we see that the mass density profiles
that we recover for the red and blue sub-samples are consistent with
each other within the error bars (though the density distribution of
the red GCs is on the higher side).

As for the deviation from sphericity is concerned, \citet{schuberth06}
suggest an ellipticity of 0.15 for NGC~4636. This is in reference to
the photometric appearance of the galaxy while observational
constraints on the spatial distribution of the dark matter halo of the
galaxy, within which the globular cluster system resides, is harder to
come by\footnotemark. \citet{schuberth06} treat NGC~4636 as spherical,
motivated by the confirmation of ``modest'' deviations from sphericity
in the inner parts of the galaxy (courtesy Dirsch et. al 2005), where
most of the GCs live. 

However, it needs to be appreciated that in CHASSIS, the assumption of
sphericity pertains to the geometry of the total gravitational
potential that the set of test particles sit in, rather than to the
spatial distribution of the test particles. Thus, we assume that it is
the outer parts of NGC~4636 (including its dark halo) that is
spherical. In any case, mistaking an ellipsoidal system as spherical
will imply overestimation of the enclosed mass profile that is
calculated from the recovered mas density distribution.  Erroneous
formulation of the potential will also add uncertainty to the recovery
of the mass density distribution itself though seeking to quantify
this bias is again not direct, given that it depends on the very
unknown that we are trying to constrain, namely the mass distribution
in this system.

\footnotetext{ The globular cluster system in M87 was reported to be
elliptical in projection \citep{mclaughlin94}, though \citet{cote01}
approximate the 3-D spatial distribution of this system as spherical.}

\subsection{Comparison with Other Work}
\noindent
The hot X-ray emitting interstellar medium (ISM) has been well-studied
\citep{jones02,loewenstein03,ewan05}, with the deepest observations
going out to about 30~kpc which is a radial extent that is well
sampled by the GCs used in this analysis.  \citet{matsushita98} and
\citet{loewenstein03} concluded that NGC 4636 is an extremely dark
matter dominated galaxy, on the basis of their analysis of ASCA and
Chandra X-ray observations respectively, by fitting hot plasma models
and assuming hydrostatic equilibrium. When compared carefully, their
mass profiles appear to be significantly different from each other -
the enclosed mass profile of \citet{matsushita98} flattens out beyond
10~kpc, reaching $5\times 10^{11}\,M_\odot$ at 30 Kpc, while the
\citet{loewenstein03} profile rises steadily to beyond
$10^{12}\,M_\odot$ at the same distance.

However, as is seen from the detailed analysis of the Chandra and
XMM-Newton observations by \citet{jones02} and \citet{ewan05}, the hot
gas has large-scale arm like features that are indicative of AGN
outburst, and thus the equilibrium assumption might not be a good one.
These observations also indicate that at about 30 kpc, there is an
abrupt transition of the hot interstellar medium of the galaxy
NGC~4636 to the intra-cluster medium of the group surrounding it,
where the gas is of significantly lower temperature and
abundance. From these observations, it is hard to decouple the dark
matter halo of the surrounding group from the dark halo of the galaxy
itself, but inner to 30~kpc, the dark halo of the galaxy must
dominate.

In terms of the shape of the predicted mass profiles, we find that
unlike the mass profiles of \citet{matsushita98}, (which exhibit a
distinct flattening outside an intermediate radius of $\sim$10 kpc),
our cumulative mass profiles do not indicate any sharp flattening -
instead, the slope of the profile  gradually tapers off from
about 10 kpc, such that even at about 50 kpc, it is only slightly
rising.

This is similar in nature to the mass profile obtained by
\citet{schuberth06} from their Model~14, i.e. their model with the
highest enclosed mass within 30 kpc. In fact, \citet{schuberth06}
speculate that the flattening observed by \citet{matsushita98} could
be a peculiarity of the ASCA data itself. In any case, the apparent
lack of hydrostatic equilibrium indicates that X-ray derived mass
profiles might not be reliable. This may explain why our estimate of
the mass of NGC 4636 falls short of what \citet{loewenstein03} suggest
on the basis of their X-ray studies.

The mass profiles inferred from a Jeans analysis of the GC kinematics
\citep{schuberth06} also yield high values of the dark matter content
in this region - the highest and lowest values from their models yield
a cumulative mass of 0.5-0.95$\times 10^{12}\,M_\odot$ at 30~kpc.  At
the error level of +1-$\sigma$, our recovered mass at about 30~kpc is
in excess of that suggested by the \citet{schuberth06} analysis, from
the same observational data, the lowest range of our errors being just
about consistent with the highest range obtained by them. Here we
recall that over-estimation of mass density, at radii where velocity
anisotropy is present, is expected to be an artifact of the assumption
of isotropy (though as discussed in Section~4.1, the effect of this
artifact is not expected to be significant). Thus, the mass-to-light
ratios presented in this paper are expected to be on the higher side
of the true values in the two bands discussed ($K$ and $R$). The
recovered velocity dispersion profile embraces the dispersion values
depicted by \citet{schuberth06}, within the $\pm$1-$\sigma$ error
bars.

It needs to be appreciated that our work is not reliant on photometry
for the extraction of the all important total mass density of the
system, and thus we are able to offer mass estimates much further into
the halo of NGC 4636, than any of the earlier attempts. We can, in
fact, calculate the mass density to just inside 7$R_e$. Of course, the
computation of the $M/L$ profiles is constrained by the limitations of
photometry.


\section{Discussion: Dark Matter in the halo of NGC 4636}
\noindent
We compute the circular velocity profile of the GC system of NGC~4636
to understand the distribution of dark matter in this system. This
$v_c$ profile is shown in black, as a function of radius in the left
panel of Figure~\ref{fig:vc}. The profile in red is given as the function
$\sqrt{GM_{\rm 3R_e}/r}$, where $G$ is the universal gravitational
constant and $M_{\rm 3R_e}$ is the mass that is enclosed within 3
effective radii. Thus, the function plotted in red merely exhibits a
Keplerian $r^{-1/2}$ fall-off with radius. The errors on this function
are the $\pm1-\sigma$ errors on $M_{\rm 3R_e}$, as recovered by
CHASSIS.

As is apparent from this figure, there is no significant difference between
the black and red profiles to about 45 kpc. This indicates that the
dark halo contribution to the mass, in the radial range 3$R_e$ to
about 5.8$R_e$ is not significantly different from the mass at
3$R_e$. This lower radius would be normally expected to bear a lower
dark matter contribution than the part of the galaxy outside
5$R_e$. That this expected trend is not statistically valid for the
case of NGC 4636, can be interpreted to conclude that the distribution
of mass in the dark halo of this galaxy is highly concentrated. This
explanation suggests itself readily, since the prospect of low dark
mass content can be ruled out on the basis of the very high $M/L$
ratio values that we have recovered.

Driven by this hunch, we proceeded to fit an NFW density distribution
\citep{nfw} to the density that is recovered by CHASSIS, from one of
the presented runs (say, RUN~I). Of course, this fit is non-trivial
and depends on the radial range in the data over which the fit is
sought. We performed a recursive routine to constrain the mass ($M_s$)
and length scales ($r_s$) in the NFW form, such that the extracted
$r_s$ lay within the radial range under consideration. The iterations
were carried out till convergence was spotted within the errors of the
fit, which were high. The radial range over which the fit was sought
extended from 4$R_e$ to about 7$R_e$ while the values of $r_s$ and
$M_s$ were recovered to be 33.7 kpc$\pm$11$\%$ and about
1.7$\times$10$^{12}$ M$_{\odot}\pm{8}\%$, respectively. This fit is
shown, superimposed in red, on the recovered density (in black), in
the right panel of Figure~\ref{fig:vc}. This recovered $r_s$ would indicate a
very high concentration of at least 9, for an assumed $M_{200}$ of at
least 2$\times$10$^{12}$ M$_\odot$. This is indicative of a strongly
concentrated system.

\citet{lintott06} have found that in a sample of about 2000
ellipticals from the Sloan Digital Sky Survey, the distribution of $c$
is log-normal, with the peak lying in the range of 3-10. Thus, on this
scale, the dark halo of NGC 4636 is predicted to be very highly
concentrated indeed!

Recently, \citet{napolitano07} have suggested a dichotomy in the dark
matter distribution of nearby early-type galaxies (probed by
the Planetary Nebula Spectrograph), where one class 
(the ``ordinary'', fast-rotating, discy/cuspy, early type
systems) exhibit rapidly dropping velocity curves while the other
slowly rotating, boxy/cored systems display flatter trends in their
rotation curves. From Figure~\ref{fig:vc} we can see that the medial
value of the circular velocity increases from about 400 km s$^{-1}$ to
an intermediate peak of about 450 km s$^{-1}$, to drop to around 400
km s$^{-1}$ again, at 5$R_e$. Thus, the rotation curve of this galaxy
does not display the kind of ``pseudo-Keplerian'' fall-off of the that
\citet{napolitano07} identify with the ``ordinary'' class of
ellipticals. Thus, our interpretation of NGC 4636 as a dark-matter
rich galaxy is consistent with the latter class of galaxies. 

\section*{Acknowledgments}
\noindent
We thank Tom Jarrett for supplying us with unpublished 2Mass
photometry for the $K$-band surface brightness profile, and Mike
Merrifield for very useful discussions. Thanks to Ylva Schuberth for
sharing observational results with us before publication. DC is
funded by a Royal Society Dorothy Hodgkin Fellowship.

\end{document}